\magnification 1100
%\magnification 1200 

\hoffset=-0.3truein
\hsize=7.1truein
\voffset=-0.4truein
\vsize=9.8truein

% First generate greater than or approximately equal to and less than or
% approximately equal to signs, using macros defined in PLAIN.TEX
\catcode`\@=11
\def\gapprox{\mathrel{\mathpalette\@versim>}}
\def\lapprox{\mathrel{\mathpalette\@versim<}}
\def\@versim#1#2{\lower2.45pt\vbox{\baselineskip0pt\lineskip0.9pt
    \ialign{$\m@th#1\hfil##\hfil$\crcr#2\crcr\sim\crcr}}}
\catcode`\@=12

\noindent .
\vskip 2cm % four lines , 2 cm
\font\sixteenbf=cmr10 at 16pt
\noindent
{\sixteenbf 
\baselineskip = 20pt % 16 pt
\emergencystretch 100pt
COMPACT X-RAY SOURCES IN NEARBY

\noindent
GALAXY NUCLEI}
\baselineskip = 12pt
\vskip 1cm % two lines
\noindent 
Edward J. M. Colbert$^{1,2}$ \& Richard F. Mushotzky$^1$
\vskip 1cm % two lines
\noindent
$^1$Lab. for High Energy Astrophysics, NASA/GSFC, Greenbelt, MD  20771

\noindent
$^2$NAS/NRC Research Associate
\vskip 1.5cm % 3 blank lines
\noindent
{\bf ABSTRACT}
\vskip 0.5cm

\noindent
We have found compact, near-nuclear
X-ray sources in 21 (54\%) of a complete sample of
39 nearby face-on spiral and elliptical galaxies 
with available ROSAT HRI data.  
%We associate these compact X-ray
%sources with accreting black holes: sources similar to active galactic
%nuclei, black hole X-ray binaries, or some new unknown type of X-ray
%source.  
ROSAT X-ray luminosities (0.2 $-$ 2.4 keV) of these compact X-ray
sources are $\sim$10$^{37}$$-$10$^{40}$ erg~s$^{-1}$.  The 
mean displacement between
the location of the compact X-ray source and the
optical photometric center of the galaxy is $\sim$390 pc.
ASCA spectra of six of the 21 galaxies show the presence of a hard
component with relatively steep ($\Gamma \approx$ 2.5) spectral slope.
A multicolor disk blackbody plus power-law 
model fits the data from the spiral galaxies
well, suggesting that the X-ray objects in these galaxies may be similar
to a black hole candidate (BHC) in its soft (high) state.  ASCA data from the
elliptical galaxies indicate that hot (kT $\approx$ 0.7 keV) gas 
dominates the emission.
The fact that
the spectral slope of the spiral galaxy sources
is steeper than in normal type 1 active galactic
nuclei (AGNs) and that
relatively low absorbing columns (N$_H \approx$ 10$^{21}$ cm$^{-2}$)
were found to the power-law component indicates that these objects are
somehow geometrically and/or physically different from AGNs in normal
active galaxies.
The X-ray sources in the spiral galaxies may be BHCs,
low-luminosity AGNs, or possibly X-ray luminous supernovae.  We estimate
the black hole masses of the X-ray sources in the spiral galaxies
(if they are BHCs or AGNs)
to be $\sim$10$^2$$-$10$^3$ M$_\odot$.
The X-ray sources in the elliptical galaxies may be BHCs, AGNs or young
X-ray supernova also.

\vskip 1cm % 2 blank lines
\noindent
{\bf INTRODUCTION}
%\vskip 0.5cm
\medskip

\noindent
An important unanswered question in extragalactic astronomy is whether
supermassive black holes (BHs) exist in the nuclei of most galaxies.
Since most galaxies are classified as ``normal,'' this question
translates into:  Are supermassive BHs present in the nuclei of most
normal galaxies?
Dynamical searches for BHs in normal galaxies reveal central
dark objects with masses $\sim$10$^6$$-$10$^{9.5}$ M$_\odot$ in many
nearby galaxies (cf. Kormendy \& Richstone 1995; Faber, these
proceedings), so it is feasible that
many normal galaxies do contain nuclear supermassive BHs.  However,
normal galaxy nuclei do not show optical evidence for broad-line
regions and narrow-line regions and 
their optical/ultraviolet continua do not typically show the signature of the 
``big blue bump,'' which is thought to be emission from the accretion
disk in AGNs.  One possible model of galaxies purports that 
supermassive BHs
are present in all galaxies, but are ``active'' in some and ``inactive''
in the others (e.g. Cavaliere \& Padovani 1988).  In this scenario, normal
galaxies could be galaxies in which the BH/accretion disk
is relatively ``inactive.''

%\vskip 0.5cm
%\noindent
%Only a few different types of point sources in external galaxies emit at
%high enough count rates so that X-ray observations can detect them: X-ray
%bright supernova remnants (cf. Schlegel 1995), luminous X-ray binaries,
%and luminous galactic nuclei.  At other wavelengths (e.g., optical),
%there are much more possible sources of emission besides compact
%galactic nuclei.  Therefore, X-ray observations are unique in their
%ability to identify accreting BHs in galactic nuclei.

\medskip
%\vskip 0.5cm
\noindent
As an example, consider the X-ray source in 
the nuclear region of the nearby normal galaxy NGC~1313.
A bright X-ray source in the nuclear region of this galaxy is
located $\sim$1 kpc from the optical nucleus, has 
L$_X(0.2-2.4~{\rm keV}) \sim$ 10$^{40}$
erg~s$^{-1}$, implying a BH mass of $\sim$10$^3$$-$10$^4$ M$_\odot$
for
an Eddington ratio of 10$^{-2}$$-$10$^{-1}$ (Colbert et al. 1995).
There is no evidence for an AGN
at wavelengths other than X-ray, yet
the implied mass is much larger than those of typical BH X-ray binaries
(XRBs)
($\sim$5$-$10 M$_\odot$, Tanaka 1989).  %\& Lewin 1995).  
%
%\vskip 0.5cm
%\noindent
We seek to understand the
nature of these near-nuclear compact X-ray sources.  Do they represent a
class of low-luminosity AGN perhaps similar to Seyfert nuclei?
%, e.g. AGNs with advection-dominated accretion flows (ADAFs)?  
Or are they BH XRBs with BH masses of $\sim$10$-$10$^4$ M$_\odot$ ?
Or are they a new class of object with properties yet to be discovered?

\medskip
%\vskip 1 cm
\noindent
{\bf SAMPLE SELECTION, OBSERVATIONS AND DATA REDUCTION}

\medskip
%\vskip 0.5cm
\noindent
%Our goal was to select a sample of galaxies in which compact
%X-ray sources could be associated with accreting BHs as unambiguously as
%possible.
All galaxies with recessional velocities smaller than 1000 km~s$^{-1}$
were selected from an electronic version (vers. 3.9) of the 
{\it Third Reference Catalog of Bright Galaxies} 
(de~Vaucouleurs et al. 1991;RC3).  
%Since X-ray luminosities are more ambiguous for sources with large absorbing 
%columns, w
We omitted galaxies
in which the disks were edge-on, since the additional large absorbing
columns can significantly reduce the soft X-ray flux of the X-ray
sources.
%make the count rates of the X-ray sources low.
% and produce low-quality spectra.  
We also omitted starburst galaxies, since they typically have multiple
sources of X-ray emission and their X-ray morphologies are generally
complex.
See Colbert \& Mushotzky (1998) for more details on the selection
criteria.
%More specifically, we required that
%the logarithm of the axial ratio be smaller than 0.4 (as taken from
%RC3).  
%Starburst galaxies produce X-ray emitting galactic winds which
%extend out of the nuclear region and the nuclear starbursts themselves
%are sources of X-ray emission (often with very complex structure), 
%so we chose to omit starburst galaxies
%from our sample.  
%We deselected starburst galaxies by requiring L$_{FIR}$
%(calculated from IRAS fluxes using the method described in Fullmer \&
%Lonsdale 1989) to be less than the blue B luminosity L$_B$.  
%We then 
%required that there be sufficiently good (exposure time $\gapprox$ 10 ks) 
%ROSAT HRI data available in the
%public archives, so that luminosities $\gapprox$ 10$^{39}$ erg~s$^{-1}$
%could be sampled. 
%The Seyfert galaxy NGC~4151 was omitted from the sample since it is
%already known to have a bright 
%X-ray emitting AGN.  The Local Group galaxies SMC and LMC
%%were also omitted from the sample since they project too large an area on
%the sky to make ROSAT imaging feasible without mosaicing.
%Galaxies in the local group were omitted? 
Our selection 
criteria produced a list of 39 nearby galaxies.
%
%\vskip 0.5cm
%\noindent
All available data for the selected 39 galaxies were retrieved from the 
US ROSAT archives at NASA/GSFC.  
%We used the IRAF/PROS software package
%to produce event list files, which were then used for imaging.  All
%imaging data were first blocked by a factor of two to give 1" pixels,
%and were then smoothed with a Gaussian profile with $\sigma =$ 3".
%Count rates from the HRI images were determined using circular source 
%regions of radius 15", centered on the peak pixel.  
%Background counts were taken from annular
%regions surrounding the source.  It was necessary in a few cases to
%restrict the azimuthal angle of the background region so that there
%would not be any contamination from nearby sources.
For the galaxies M33, M81, NGC~4374, NGC~4406 and NGC~4552, diffuse X-ray
emission was noticed to be present in the nuclear region.
%, i.e., in the circular source regions.
For these galaxies, we modelled the radial profile of the X-ray emission
to estimate the luminosity of the point-like component.

\medskip
%\vskip 0.5cm
\noindent
Data for eight observations (of six galaxies) in our sample were
retrieved from the ASCA archives.
%in order to facilitate our
%understanding of the nature of the compact X-ray sources.  
The six
galaxies selected were three spiral galaxies (M33 [two observations], 
NGC~1313 [two observations], and NGC~5408) and three elliptical galaxies
(NGC~4374, NGC~4406 and NGC~4552).  These galaxies
were selected because (1) they have compact X-ray sources in
the HRI images, (2) no adjacent X-ray sources were present in the
ASCA fields (so that the ASCA spectra include emission only from the
compact X-ray source under study), and (3) public archival ASCA data 
were available.  
%The raw (``unscreened'') ASCA data were screened using the XSELECT
%software program (part of the FTOOLS X-ray analysis software package)
%using standard screening restrictions (see Colbert \& Mushotzky 1998).

\medskip
%\vskip 1 cm
\noindent
{\bf RESULTS FROM ROSAT HRI SURVEY}

\medskip
%\noindent
%\underbar{Overview of Results from HRI Survey}
%\medskip

\noindent
Of the 39 galaxies for which HRI imaging data were available, we found 
(detection $>$ 3$\sigma$)
compact X-ray sources within $\sim$1' of the nucleus of 21
galaxies.  Lira et al. (these proceedings) find a similar result.
Here we have used the optical position listed in RC3 (the 
photometric center) as the location of the nucleus. X-ray fluxes and 
luminosities in the 0.2$-$2.4 keV band were calculated from the count
rates, assuming a power-law (photon index $\Gamma =$ 1.7) spectrum 
and absorption from the Galactic hydrogen column (Dickey \& Lockman
1990).

\medskip
%\noindent
%\underbar{X-ray Luminosities}
%\medskip

\noindent 
A histogram of the 0.2$-$2.4 keV luminosities is shown in Figure 1.
Luminosities of the compact sources ranged from 8 $\times$10$^{36}$
erg~s$^{-1}$ to $\sim$10$^{40}$ erg~s$^{-1}$, with a mean of 
3 $\times$ 10$^{39}$ erg~s$^{-1}$.
For reference, Seyfert galaxies typically
have L$_X \sim$ 10$^{42}$$-$10$^{44}$ erg~s$^{-1}$ and the Eddington
luminosity of a 1.4 M$_\odot$ accreting neutron star is 10$^{38.3}$
erg~s$^{-1}$.
So, on average, the X-ray sources are at least one order of magnitude
more luminous than the most luminous (i.e., Eddington ratio of unity) 
neutron star X-ray binaries, but are several orders of magnitude less
luminous than typical AGN in Seyfert nuclei.  
Also shown in Figure 1 are the galaxy types of the host galaxies.  Note
that the most luminous X-ray sources (L$_X \sim$ 10$^{40}$
erg~s$^{-1}$) tend to be found in elliptical hosts.

\medskip
%\noindent
%\underbar{Separation from Galaxy Center}
%\medskip

\noindent
The offsets of the X-ray
sources from the galaxy nucleus (photometric center) is significant in
many of the sources.  
An interesting result of our study is that $\sim$40\% (9 of 21)
of the compact sources were located $\gapprox$ 10" 
($\gapprox$ 0.3 kpc at the mean distance of our sample)
from the
photometric optical center (as listed in RC3).  For comparison, most
observed objects have ROSAT X-ray positional errors of $\lapprox$10"
(Briel et al. 1996).
In Figure 2, we show a plot of L$_X$ vs. the angular separation between
the X-ray source and the photometric center.  It is noteworthy that the
mean luminosity of sources with small ($\lapprox$200$-$400 pc)
separation is similar to the mean luminosity of sources with 
large ($\gapprox$200$-$400 pc) separation.

%\vskip 1 cm
\medskip
\noindent
{\bf RESULTS FROM ASCA SPECTRAL FITTING}

\medskip
\noindent
We first tried fitting the spectra with two single component models: (1)
a Raymond-Smith (RS) thermal plasma and a power-law (PL) model.  The
power-law was, in general, a better fit for the spiral galaxy 
spectra, while the RS model was a better fit to the elliptical galaxy spectra.
The spiral galaxies were fit

\vfill \eject
\noindent .
\vskip 3.5in
\noindent
{\bf Figures:} Figure 1 (left) shows a histogram of the 0.2$-$2.4 keV
luminosities and the galaxy ``T'' type as listed in RC3.  Figure 2 (right)
shows the 0.2$-$2.4 keV luminosity {\it vs.} the separation (Delta)
between the position of the X-ray source and the photometric center of
the galaxy (from RC3).  Unclassified galaxies are shown as open
rectangles and active galaxies are shown as filled rectangles.

\medskip

\noindent
with a relatively steep power law slope 
($\Gamma =$ 2.3 $-$ 2.8), except for one of the observations of
NGC~1313,
 for which $\Gamma =$ 1.7.
We also tried modelling the data with multiple component fits, for
example a RS+PL model, and a disk blackbody (DBB, Mitsuda et al. 1984)
plus power-law model (DBB+PL).  In nearly all cases, the two-component
fits are preferred.
% with very low probabilities of exceeding the F values
%by chance (significance of the extra terms $\gapprox$ 97$-$99.9\%).  
%An exception is M33, in which the single component RS model is nearly
%equally well fit as the DBB+PL fit.
%
%\medskip
%\noindent
It is worthwhile to note that the elliptical galaxies are much better
fit by a RS+PL model compared to a DBB+PL model, indicative of a hot gas
component which is known to be common in these galaxies.  
%We also tried
%a three-component model consisting of a RS thermal plasma, a DBB, and a
%power-law model.  These fits are much more significant fits than the
%DBB+PL fits, but are not significantly preferred over the RS+PL models.
%Thus, we conclude that a disk blackbody component is not required by the
%spectral data and the RS+PL fits are the best fit models for the
%elliptical galaxies in our sample.

\medskip
\noindent
We next discuss the results of the spiral galaxy spectra in terms of
emission expected from a BH XRB.  In the hard state, BH XRBs typically
have hard spectra with power-law slopes $\Gamma \approx$ 1.3 $-$ 1.9,
whereas in the soft state, the power-law slope steepens to $\approx$ 2.4
(e.g., Ebisawa, Titarchuk, \& Chakrabarti 1996) and a soft component
appears in the X-ray spectrum (usually modelled as a multicolor disk
blackbody component).  One of the observations of NGC~1313 exhibits a
slope more consistent with a BH XRB in its hard state, while in the other 
observation (2.4 years later) the slope steepens to $>$2.5, more
consistent with a soft state.
To further investigate the hypothesis that the emission is from XRB-like
objects in their soft state, we fitted the spectra from these galaxies
with the bulk-motion (BM) model promoted by Titarchuk and collaborators
(e.g., see Shrader \& Titarchuk 1998).  In general, the fits were
acceptable, although the reduced $\chi^2$ values were typically higher
than those for the RS+PL and DBB+PL fits.  Again, the power-law slopes
were quite steep ($\sim$2.5, except for the later observation of NGC~1313).
The normalization to the BM model fits allows us to directly estimate
the
masses of the central BH.  We estimate M$_{BH}$ values of 117, 590, and 
$\sim$8500 M$_\odot$ for M33, NGC~1313, and NGC~5408, respectively.

\medskip
%\vskip 1 cm
\noindent
{\bf DISCUSSION}
%\vskip 0.5cm
\medskip

\noindent
For both the fits to the elliptical galaxies and the spiral galaxies,
the power-law slopes are considerably steeper than found for ASCA
analysis of type 1 AGN, in which $<\Gamma> =$ 1.91 $\pm$ 0.07 (Nandra et
al. 1997).  The mean value and standard deviation for our three
elliptical galaxies is 2.55 $\pm$ 0.76.  This may be indicative of a
different kind of emission process, for example both Narrow-Line
Seyfert 1 galaxies and BH XRBs in their soft state have comparably steep
power-law slopes.  The power-law slopes for the spiral galaxies are also
comparatively steep.  The mean and standard deviation of the slope $\Gamma$ 
(excluding the single observation of NGC~1313 in which the slope was shallow)
is 2.54 $\pm$ 0.38 (DBB+PL fit).

\medskip
\noindent
There does not seem to be a single explanation that works for all of the
compact X-ray sources.  For example, the X-ray luminosities of the
sources in the galaxies with evidence for nuclear activity in the
optical/UV are
%significantly 
larger than those luminosities of the sources in the
otherwise normal (or ``unclassified'') galaxies (mean 0.2$-$2.4 keV
luminosity 4.4 $\times$ 10$^{39}$ erg~s$^{-1}$ {\it vs.} 2.4 $\times$
10$^{39}$ erg~s$^{-1}$).
The mean separation of the X-ray sources
from the galaxy nucleus is also larger for the normal galaxies (see Figure 2).
This suggests that there may be two different types of X-ray emitting objects
in these two types of galaxies.
Compact X-ray sources are found in nearly all cases in which previous
possible evidence for a BH or an AGN had been found (e.g., large-scale
radio sources, dynamical dark core masses $>$ 10$^6$ M$_\odot$, or
optical spectral classification as Seyfert, LINER, or ``transition''
object).  This indicates that some of the X-ray sources in our sample
are probably AGN or at least accretion-driven objects with
supermassive BHs.

\medskip
\noindent
The soft components in the ASCA spectra in the elliptical galaxies are
better fit by a Raymond-Smith thermal plasma than a multicolor disk
blackbody model.  This suggests that some of the X-ray emission from the
central X-ray sources in the elliptical galaxies may be emission from
hot gas, a component which has been known to be present in many other
elliptical galaxies (e.g., Matsushita et al. 1994).
%
%We analyzed archival ASCA spectra of six of the 21 galaxies with
%compact X-ray sources (3 spiral galaxies and 3 elliptical galaxies).
The elliptical galaxy spectra are adequately fit with a two-component
Raymond-Smith thermal plasma model plus an absorbed power-law model.
%The intrinsic absorption of the power-law source was typically several
%$\times$ 10$^{21}$ cm$^{-2}$ (significantly smaller than seen in Seyfert
%2 galaxies) and the power-law slope was significantly steeper ($\Gamma
%\approx$ 2.5) than seen in type 1 or 2 AGNs.  
The spiral galaxy spectra
are well-fit by a two-component model consisting of a multicolor disk
blackbody model plus a power-law model.  Assuming both components 
suffer the same intrinsic absorption, we find intrinsic absorption of
a few 10$^{21}$ cm$^{-2}$ and, again, steep power-law slopes of 
$\Gamma \approx$ 2.5.  
%The central X-ray source in the spiral galaxy 
%NGC~1313 seems to have changed from a hard state to a soft state in the
%2.4 intervening years between observations.  
%The spiral galaxy X-ray
%sources, when fit with the Bulk Motion Comptonization model of Shrader
%\& Titarchuk (1998), predict black hole masses $\sim$10$^2$$-$10$^3$
%M$_\odot$.

\medskip
\noindent
What does seem to be universal among the X-ray sources is that they have
relatively steep spectral slopes compared with those of Seyfert 1 nuclei.
This, and the fact that large absorbing columns are not found, suggests
that the X-ray sources are probably different physically or
geometrically from typical AGN in active X-ray bright galaxies.

%We should also mention that the preference for low absorbing columns
%could also be due to a selection effect.  X-ray source with similar
%intrinsic (unabsorbed) luminosities but larger absorbing columns (e.g.,
%N$_H$ $\sim$ 10$^{23}$$-$10$^{24}$ cm$^{-2}$) would be considerably
%weaker X-ray sources and therefore may have led ASCA and ROSAT observers
%not to observe those objects.  
%%Any such objects that were observed would have lower count rates and may 
%% be too weak to be analyzed properly.  Not so for our sample because we
%% selected objects based on exposure time.

%Using archival ROSAT HRI observations, we have used a distance-limited sample
%of 39 nearby galaxies to search for compact ``nuclear'' X-ray sources.
%We found compact X-ray sources within $\sim$1' of the nucleus
%(photometric center) of 21 of the 39 galaxies.  Approximately 40\% (9 of
%21) of the compact X-ray sources were located $>$10" ($\gapprox$ 0.3 kpc
%at the mean distance of our sample) from the photometric center of the
%galaxy.  

\medskip
\noindent
{\bf CONCLUSIONS}
\medskip

\noindent
The emerging picture of the compact X-ray sources in nearby galaxies is
that they are probably high-mass (M $\gapprox$ 100 M$_\odot$) 
accreting
black holes (or X-ray luminous supernovae with no historical optical
evidence).
It is not clear how black holes with mass $\sim$10$^2$$-$10$^4$
M$_\odot$ might form.
It is not clear whether they are binary systems or not,
but the steep hard X-ray emission is consistent with that of a BHC in its
soft (high) state.  Such steep slopes are also consistent with what has
been observed in Narrow-Line Seyfert 1 galaxies.  
%The elliptical
%galaxies may have compact X-ray sources that are due to peaks in
%the distribution of hot gas in the galaxy nucleus.

%\vskip 1 cm
\medskip
\noindent
{\bf ACKNOWLEDGEMENTS}

%\vskip 0.5cm
\medskip
\noindent
EJMC thanks the National Research Council for financial support and
Lev Titarchuk for many enlightening discussions.

\medskip
%\vskip 1cm
\noindent 
{\bf REFERENCES}
\medskip

\def\rf{\hang\noindent}
\def\apj{ApJ}

%\rf Yummy, Y., Inmytummy, N., And, A., More, B., People, C., Toadd, D.,
%    Tothis, E., List, F., Plus, G., Afew, H., More, I. 1997, 
%    Yesimhungry, 49, 94

%\rf Antonucci, R. R. J. 1993, ARA\&A, 31, 473

%\rf Beckman, J., Cepa, J., Prieto, M., \& Tunon, C. M. 1987, Rev. Mex.
%   Astron. Astrof., 14, 134

%\rf Belloni, T., Mendez, M., King, A. R., van der Klis, M., \& van
%   Paradijs, J. 1997, \apj, 488, L109

%\rf Bender, R., Kormendy, J., \& Dehnen, W. 1996, \apj, 464, L123

%\rf Bothun, G. D. 1986, AJ, 91, 507

%\rf Bridle, A. H., \& Perley, R. A. 1984, ARA\&A, 22, 319

\rf Briel, U. G., et al. 1996, ROSAT User's Handbook (December 1996 HTML
  test version) % --
% http://ftp.rosat.mpe-garching.mpg.de/rosat\_svc/doc/handbook/html)

%\rf Brown, G. E., \& Bethe, H. A. 1994, \apj, 423, 659

\rf Cavaliere, A., \& Padovani, P. 1988, \apj, 333, L33

\rf Colbert, E. J. M., \& Mushotzky, R. F. 1998, in preparation

\rf Colbert, E. J. M., Petre, R., Schlegel, E. M., \& Ryder, S. D.  1995, 
  \apj, 446, 177

%\rf Dekel, A., \& Silk, J. 1986, \apj, 303, 39

\rf Dickey, J. M., \& Lockman, F. J. 1990, ARA\&A, 28, 215

\rf de Vaucouleurs, G., de Vaucouleurs, A., Corwin, H. G., Buta, R. J.,
    Paturel, G., \& Fouque, P. 1991, Third Reference Catalog of Bright
    Galaxies (RC3; New York: Springer-Verlag)

\rf Ebisawa, K., Titarchuk, L., \& Chakrabarti, S. K. 1996, PASJ, 48, 59

%\rf Eskridge, P. B., White, R. E., III, \& Davis, D. S. 1996, \apj, 463, L59

%\rf Ford, H. C., et al. 1994, \apj, 435, L27

%\rf Forman, W., Jones, C., \& Tucker, W. 1985, \apj, 293, 102

%\rf Fullmer, L., \& Lonsdale, C. 1989, Cataloged Galaxies and Quasars in the
%      IRAS Survey (JPL Pub. D-1932, Version 2, Appendix B)
%
%\rf Harms, R. J., et al. 1994, \apj, 435, L35

%\rf Heckman, T. M. 1980, A\&A, 87, 152

%\rf Ho, L. C., Filippenko, A. V., \& Sargent, W. L. W. 1995, ApJS, 98, 477

%\rf Hummel, E., Sancisi, R., \& Ekers, R. D. 1984, A\&A, 133, 1

%\rf Johansson, L. 1987, A\&A, 182, 179

%\rf Kinney, A. L., Bohlin, R. C., Calzetti, D., Panagia, N., \& Wyse, R. F. G.
%   1993, ApJS, 86, 5 

\rf Kormendy, J., \& Richstone, D. 1995, ARA\&A, 33, 581 

%\rf Lamb, S. A., Gallagher, J. S., Hjellming, M. S., \& Hunter, D. A. 1985
%    \apj, 291, 63 

%\rf Lasota, J.-P., Abramowicz, M. A., Chen, X., Krolik, J., Narayan, R., \& 
%   Yi, I. 1996, \apj, 462, 142

%\rf Loewenstein, M., Hayashida, K., Toneri, T., \& Davis, D. S. 1998, \apj, 
%   in press (April)

%\rf Mahadevan, R. 1997, \apj, 477, 585

\rf Matsushita, K., et al. 1994, \apj, 436, L41

%\rf Melisse, J. P. M., \& Israel, F. P. 1994, A\&A, 285, 51

\rf Mitsuda, K. et al. 1984, PASJ, 36, 741

%\rf Miyoshi, M., Moran, J., Herrnstein, J., Greenhill, L., Nakal, N.,
%   Diamond, P., \& Inoue, M., 1995, Nature, 373, 127

%\rf Moshir, et al. 1990, IRAS Faint Source Catalog, version 2

\rf Nandra, K., George, I. M., Mushotzky, R. F., Turner, T. J., \&
    Yaqoob, T. 1997, \apj, 477, 602

%\rf Narayan, R. 1997, in Accretion Phenomena and Related Outflows (IAU
%   Colloq. 163, ASP Conf. Ser. Vol. 121), eds. D. Wickramasinghe, L.
%   Ferrario, \& G. Bicknell, p.~75

%\rf Narayan, R., \& Yi, I. 1994, \apj, 428, L13

%\rf Narayan, R., \& Yi, I. 1995a, \apj, 444, 231

%\rf Narayan, R., \& Yi, I. 1995b, \apj, 452, 710 

%\rf Okada, K., Dotani, T., Makishima, K., Mitsuda, K., \& Mihara, T.  1998, 
%    PASJ, 50, 25

%\rf Okada, K., Mihara, T., Makishima, K., \& The ASCA Team  1994, in 
%   New Horizon of X-ray Astronomy (Universal Academy Press) (eds.
%   Makino, F., \& Ohashi, T.), p.~515

%\rf Osterbrock, D. E. 1989, in Astrophysics of Gaseous Nebulae and Active
%  Galactic Nuclei (Mill Valley: University Science), p.~X

%\rf Petre, R., Mushotzky, R. F., Serlemitsos, P. J., Jahoda, K., \&
%  Marshall, F. E. 1993, \apj, 418, 644

%\rf Raymond, J. C., \& Smith, B. W. 1977, ApJS, 35, 419

%\rf Sandage, A., \& Hoffman, G. L. 1991, \apj, 379, L45

%\rf Schlegel, E. M. 1995, Rep. Prog. Phys., 58, 1375

%\rf Shakura, N. J., \& Sunyaev, R. A. 1973, A\&A, 24, 337

\rf Shrader, C., \& Titarchuk, L. 1998, \apj, 499, L31

%\rf Silk, J., \& Rees, M. J. 1998, A\&A, 331, L1

%\rf Takano, M., Mitsuda, K., Fukazawa, Y., \& Nagase, F. 1994, \apj, 436, L47

\rf Tanaka, Y. 1989, in Two Topics in X-ray Astronomy (23rd ESLAB
   Symp.), eds. J. Hunt \& B. Battrick (Paris: ESA), p.~3

%\rf Tanaka, Y., \& Lewin, W. H. G. 1995, in X-ray Binaries (ed. Lewin,
%   Paradijs, \& van den Heuvel), Ch. 3

%\rf Thornley, M. D. 1996, \apj, 469, L45

%\rf Tully, R. B. 1988, Nearby Galaxies Catalog (Cambridge: Cambridge
%   Univ. Press)

%\rf Turner, T. J., George, I. M., Nandra, K., \& Mushotzky, R. F. 1997, ApJS,
%   113, 23

%\rf Wrobel, J. M., \& Heeschen, D. S. 1984, \apj, 287, 41

%\rf Yaqoob, T., Ebisawa, K., \& Mitsuda, K. 1993, MNRAS, 264, 411

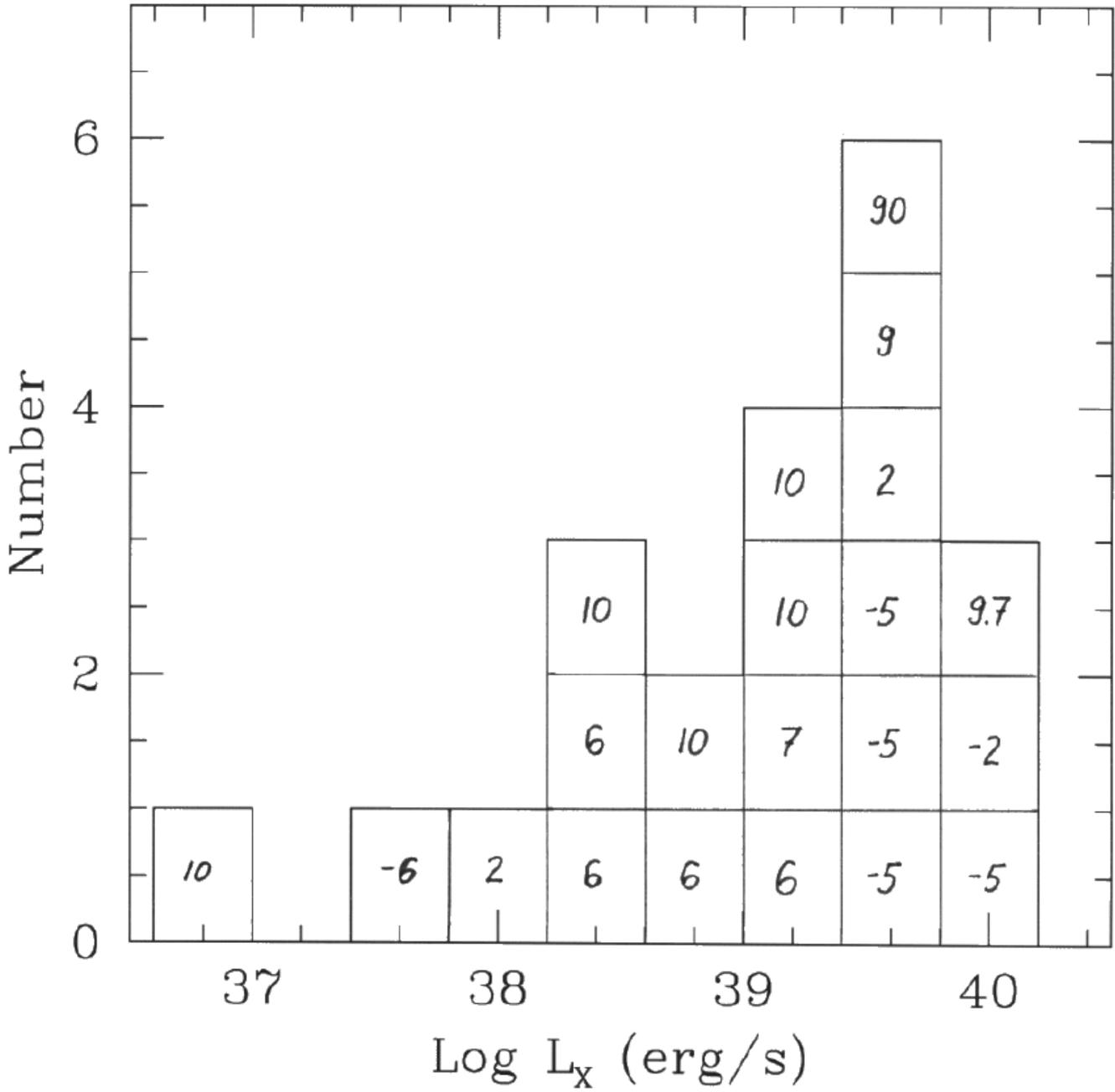
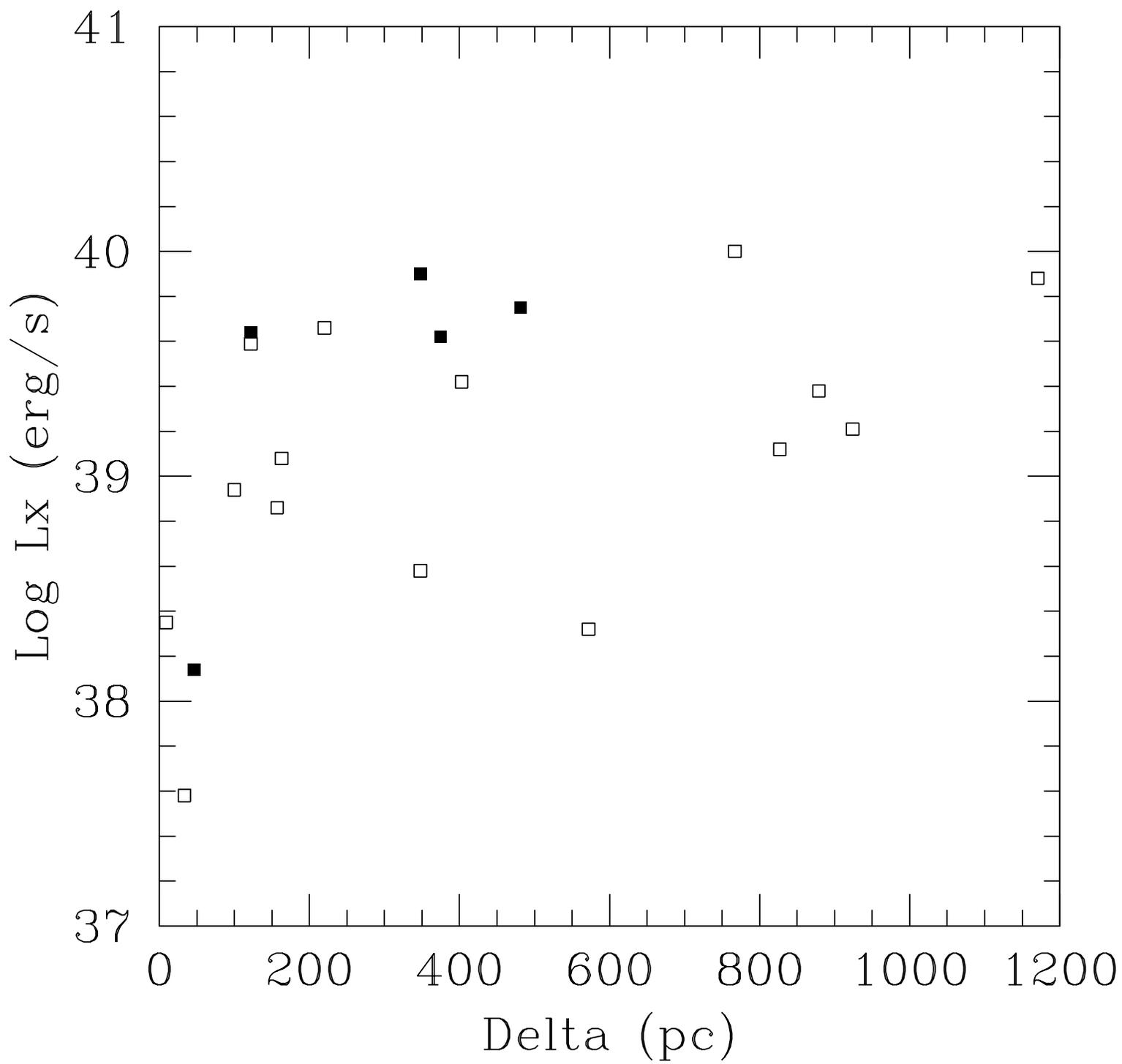
\bye